%% file: main.tex
\documentclass[lettersize,journal]{IEEEtran}
\usepackage{amsmath,amsfonts,amssymb}
\usepackage{algorithmic}
\usepackage{array}
\usepackage[caption=false,font=normalsize,labelfont=sf,textfont=sf]{subfig}
\usepackage{textcomp}
\usepackage{stfloats}
\usepackage[hyphens]{url}
\usepackage{verbatim}
\usepackage{graphicx}
\usepackage{cite}
\usepackage{enumitem}
\usepackage{xcolor}
\usepackage{relsize}
\usepackage{makecell}
\usepackage{tabularx}
\usepackage{hhline}
\usepackage{multirow}
\usepackage{diagbox}
\usepackage{adjustbox}
\usepackage[linesnumbered]{algorithm2e}
\usepackage{tikz}
\usepackage{setspace}
\def\BibTeX{{\rm B\kern-.05em{\sc i\kern-.025em b}\kern-.08em
    T\kern-.1667em\lower.7ex\hbox{E}\kern-.125emX}}
\usepackage{balance}

\def\circnumfont{\smaller\bfseries}
\def\circnumbase{99}
\newlength\circnumwidth
\newlength\circraise
\newlength\circsep
\def\circbgcolor{black}
\def\circfgcolor{white}

\makeatletter
\newcommand*\circnum[1]{% 		
  		\raisebox{\circraise}{%
  			\tikz\node[%
  				circle,draw=\circbgcolor,thin,inner sep=.25\circsep,%  				
  				top color=\circbgcolor,bottom color=\circbgcolor,%
					text width=\circnumwidth,%
					font=\circnumfont,text badly centered,\circfgcolor%
					]{#1};}}
\makeatother

\begin{document}

\title{Exploration of Systolic-Vector Architecture with Resource Scheduling for Dynamic ML Workloads}

\author{Jung-Hoon Kim$^*$, \IEEEmembership{Graduate Student Member,~IEEE,} Sungyeob Yoo$^*$, \IEEEmembership{Graduate Student Member,~IEEE,}\\
Seungjae Moon, \IEEEmembership{Graduate Student Member,~IEEE,} and Joo-Young Kim, \IEEEmembership{Senior Member,~IEEE}}

\markboth{Journal of \LaTeX\ Class Files,~Vol.~14, No.~8, August~2021}%
{Shell \MakeLowercase{\textit{et al.}}: A Sample Article Using IEEEtran.cls for IEEE Journals}

\IEEEpubid{0000--0000/00\$00.00~\copyright~2021 IEEE}

\maketitle
\def\thefootnote{*}\footnotetext{These authors contributed equally to this work}\def\thefootnote{\arabic{footnote}}

\begin{abstract}
As artificial intelligence (AI) and machine learning (ML) technologies disrupt a wide range of industries, cloud datacenters face ever-increasing demand in inference workloads.
However, conventional CPU-based servers cannot handle excessive computational requirements of deep neural network (DNN) models, while GPU-based servers suffer from huge power consumption and high operating cost.

In this paper, we present a scalable systolic-vector architecture that can cope with dynamically changing DNN workloads in cloud datacenters. 
We first devise a lightweight DNN model description format called unified model format (UMF) that enables general model representation and fast decoding in hardware accelerator. 
Based on this model format, we propose a heterogeneous architecture that features a load balancer that performs a high-level workload distribution and multiple systolic-vector clusters, in which each cluster consists of a programmable scheduler, throughput-oriented systolic arrays, and function-oriented vector processors. We also propose a heterogeneity-aware scheduling algorithm that enables concurrent execution of multiple DNN workloads while maximizing heterogeneous hardware utilization based on computation and memory access time estimation.
Finally, we build an architecture simulation framework based on actual synthesis and place-and-route implementation results and conduct design space exploration for the proposed architecture.
As a result, the proposed systolic-vector architecture achieves 10.9$\times$ higher throughput performance and 30.17$\times$ higher energy efficiency than a compatible GPU on realistic ML workloads. The proposed heterogeneity-aware scheduling algorithm improves the throughput and energy efficiency by 81\% and 20\%, respectively, compared to a standard round-robin scheduling. 
\end{abstract}


\input{Outline/1_Introduction}

\input{Outline/2_Background_and_Motivation}
\input{Outline/3_Unified_Model_Format}
\input{Outline/4_Heterogeneous_Vector_Array_Architecture}
\input{Outline/5_Scheduling_Methods}
\input{Outline/6_Evaluation_and_Result}
\input{Outline/7_Related_Work}
\input{Outline/8_Conclusion}

\newpage

\section*{Biography}

\begin{IEEEbiography}
[{\includegraphics[width=1in,height=1.25in,clip,keepaspectratio]{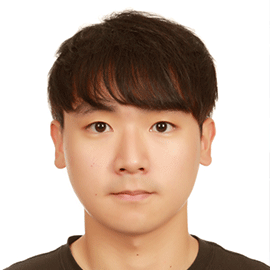}}]{Jung-Hoon Kim}
received the B.S. degree in the department of electronic engineering from Hanyang University, Seoul, South Korea, in 2020, and M.S. degree in Electrical Engineering from Korea Advanced Institute of Science and Technology (KAIST), in 2022. He is currently pursuing the Ph. D. degree at the Korea Advanced Institute of Science and Technology (KAIST), Daejeon, South Korea. His current research interests include computer architecture and ASIC chip design for machine learning.

\end{IEEEbiography}

\begin{IEEEbiography}
[{\includegraphics[width=1in,height=1.25in,clip,keepaspectratio]{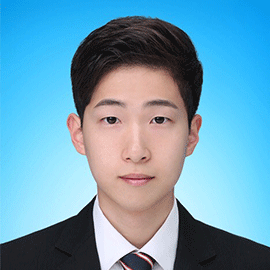}}]{Sungyeob Yoo}
received the B.S. degree in the department of electronic engineering from Ajou University, Suwon, South Korea, in 2020, and M.S. degree in Electrical Engineering from Korea Advanced Institute of Science and Technology (KAIST), in 2022. He is currently pursuing the Ph. D. degree at the Korea Advanced Institute of Science and Technology (KAIST), Daejeon, South Korea. His current research interests include computer architecture and VLSI design for datacenter.

\end{IEEEbiography}

\begin{IEEEbiography}
[{\includegraphics[width=1in,height=1.25in,clip,keepaspectratio]{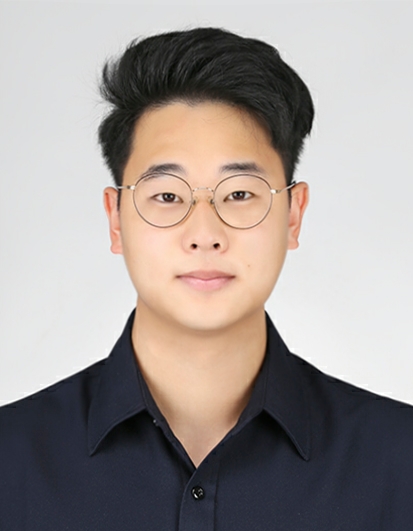}}]{Seungjae Moon}
received the B.S. in Electrical Engineering from the University of Washington, Seattle, WA, in 2020. He is currently pursuing the M.S. degree in Electrical Engineering at Korea Advanced Institute of Science and Technology (KAIST), Daejeon, South Korea. His current research interests include computer architecture, SoC design, and hardware accelerator for machine learning and other emerging applications in datacenters.

\end{IEEEbiography}

\begin{IEEEbiography}
[{\includegraphics[width=1in,height=1.25in,clip,keepaspectratio]{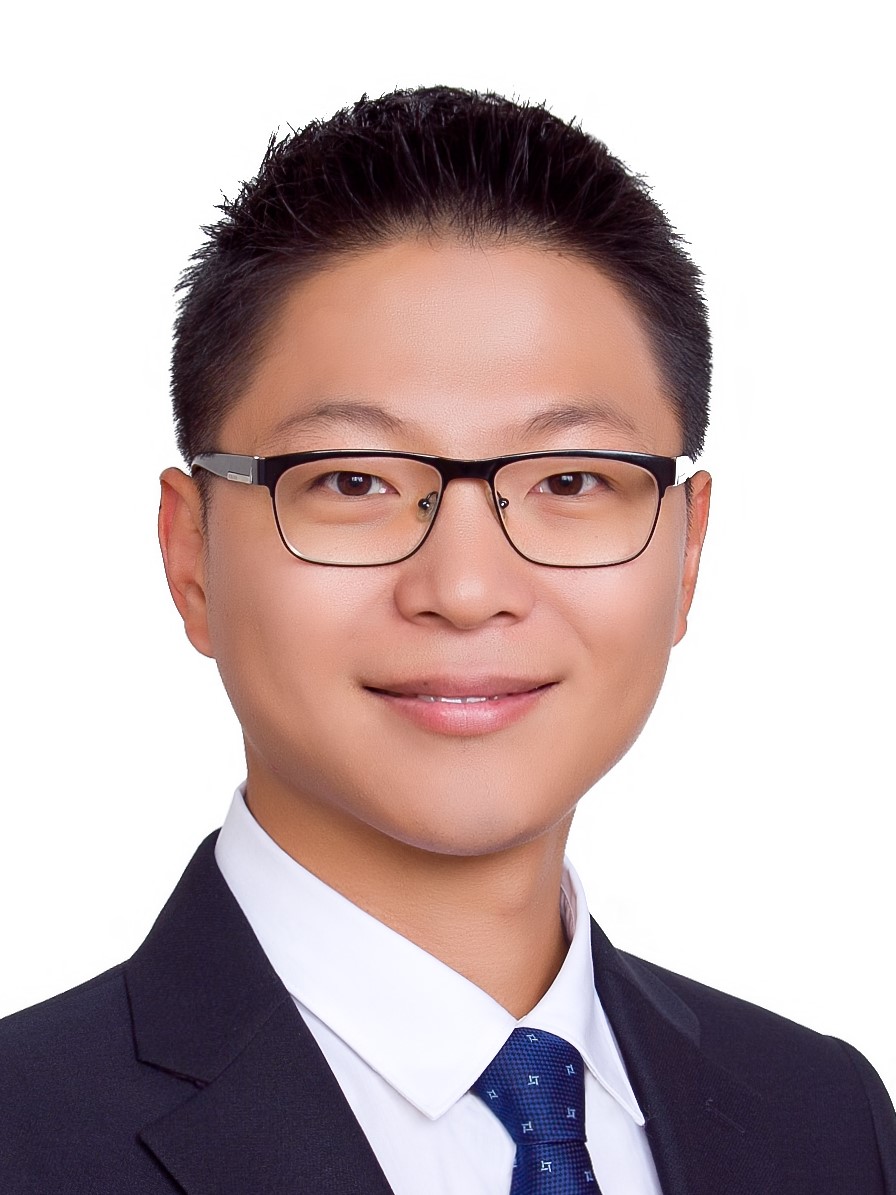}}]{Joo-Young Kim}
(M'05--SM'19)
received the B.S., M.S., and Ph. D. degrees in Electrical Engineering from Korea Advanced Institute of Science and Technology (KAIST), Daejeon, South Korea, in 2005, 2007, and 2010, respectively. He is currently an Assistant Professor in the School of Electrical Engineering at KAIST. He is also the Director of AI Semiconductor Systems Research Center. His research interests span various aspects of hardware design, including VLSI design, computer architecture, FPGA, domain-specific accelerators, hardware/software co-design, and agile hardware development. Before joining KAIST, he was a Senior Hardware Engineering Lead at Microsoft Azure, Redmond, WA, USA, working on hardware acceleration for its hyper-scale big data analytics platform named Azure Data Lake. He was also one of the initial members of Catapult project at Microsoft Research, Redmond, WA, USA, where he deployed a fabric of field-programmable gate arrays (FPGAs) in datacenters to accelerate critical cloud services, such as machine learning, data storage, and networking. 

Dr. Kim was a recipient of the 2016 IEEE Micro Top Picks Award, the 2014 IEEE Micro Top Picks Award, the 2010 DAC/ISSCC Student Design Contest Award, the 2008 DAC/ISSCC Student Design Contest Award, and the 2006 A-SSCC Student Design Contest Award. He currently serves as an Associate Editor for the IEEE TRANSACTIONS ON CIRCUITS AND SYSTEMS I: REGULAR PAPERS.
\end{IEEEbiography}

\vfill

\end{document}

%% file: Outline/1_Introduction.tex
\section{Introduction}
\IEEEPARstart{A}{rtificial} intelligence (AI) and machine learning (ML) technology change a wide range of industries, including information technology, mobile communication\cite{wang2019edge}, automotive\cite{grigorescu2020survey, ma2020cooperative}, and manufacturing\cite{patel2018raw, ghahramani2020ai}. As major cognitive applications such as computer vision\cite{simonyan2014very,sandler2018mobilenetv2,he2016deep,krizhevsky2012imagenet, iandola2016squeezenet}, speech recognition\cite{amodei2016deep,conneau2020unsupervised, baevski2020wav2vec, wang2020fairseq}, and natural language processing\cite{radford2019language, devlin2018bert} adopt ML, datacenters face ever-increasing demand in ML inference\cite{park2018deep}. In addition, deep neural network (DNN) models have been developing quickly, becoming more expensive and diversified to attain high accuracy \cite{krizhevsky2012imagenet, simonyan2014very, he2016deep, sandler2018mobilenetv2, devlin2018bert, radford2019language}.
Therefore, the computing platform in datacenters is required to address computationally intensive ML workloads generated by a number of users. More specifically, the individual hardware needs to handle the concurrent execution of ML workloads, which involves various DNN models and are dynamically changing.
However, conventional CPU and GPU based platforms are not suitable for these new ML workloads in datacenters. CPU cannot cope with the tremendous amount of computations in ML workloads, while GPU requires large power consumption with high operating costs. 
\IEEEpubidadjcol

Many domain-specific hardware accelerators\cite{chen2016eyeriss, lee2018unpu, ghodrati2020planaria, kwon2018maeri, kang2021ganpu} have been proposed to perform ML inference on a single model at high hardware efficiency. These specialized DNN accelerators use a customized dataflow to compute massive data with high processing element (PE) utilization and low data movement. However, each DNN model has a different compute-to-memory bandwidth ratio, so significant underutilization of PE or memory bandwidth occurs with DNN models that do not fit the given dataflow.
Therefore, accelerators with reconfigurable dataflow \cite{ghodrati2020planaria} and heterogeneous dataflow \cite{kwon2018maeri, kang2021ganpu} have recently been proposed. The accelerator with reconfigurable dataflow accommodates each model’s computational characteristic, thus maintaining high PE utilization for various DNN workloads, but the area and power overhead for reconfiguration are inevitable regardless of the interconnect optimization. 
The accelerator with heterogeneous dataflow distributes each workload's layer and sub-layer tasks to the optimized dataflow for efficient computation, but such accelerator is prone to low hardware utilization when a particular series of tasks continuously map to the same dataflow.
Most importantly, most of the previous accelerators focused on the execution of a single ML workload at a time, even though it may involve multiple DNN models. However, a new type of hardware accelerator that can optimize its efficiency over the multiple ML workloads in flight, in which each workload includes at least a DNN model, is necessary in the datacenter.

In this paper, we propose a scalable systolic-vector accelerator architecture that can efficiently execute multiple ML inference workloads in datacenters. Unlike previous accelerator architectures are somewhat dedicated around centered systolic arrays to maximize the throughput performance~\cite{jouppi2021ten, sharma2018bit}, the proposed architecture includes multiple systolic-vector clusters, in which each cluster includes independent systolic arrays and vector processors, as well as a programmable microprocessor for runtime resource scheduling.
Moreover, the vector processor supports the computation of the systolic array when a sequence of tasks monopolize the systolic array. This feature increases the peak performance when the systolic array is being fully utilized and allows for more diverse scheduling. The scheduler can be programmed with various scheduling algorithms for the efficient execution of ML inference workloads. In particular, we propose a heterogeneity-aware scheduling algorithm that can maximize the cluster hardware utilization. Lastly, we suggest a hardware-amenable model description packet format named unified model format (UMF) to enable both the description of the numerous DNN models from users and fast decoding in hardware.

To evaluate the proposed architecture and scheduling algorithms on realistic ML workloads in datacenters, we build a simulation framework and generate benchmark workloads imitating datacenter workloads based on a set of popular convolutional neural networks and transformer-based models. Also, we create the ONNX-to-UMF converter that extracts the essential information of each DNN request and generates UMF packets. With these features, we achieve 10.92$\times$ higher performance and 30.17$\times$ higher energy efficiency, on average, than a comparable GPU for realistic datacenter workloads.

The main contributions of our work are as follows.
\begin{itemize}[leftmargin=*]
\item {\textbf{Hardware-amenable Lightweight DNN Model Description:}} We create a lightweight DNN model description called unified model format (UMF), which enables efficient decoding in the AI accelerator.
\item {\textbf{Scalable Systolic-Vector Architecture:}} We propose a heterogeneous systolic-vector architecture, a novel DNN accelerator architecture that features a top-level load balancer and systolic-vector clusters. Each cluster is made of throughput-oriented systolic arrays and function-oriented vector processors and a scheduler that allocates the workload to the underlying processors at runtime to serve dynamically changing DNN workloads in datacenters.
\item {\textbf{Hardware Resource Maximization:}} We devise a heterogeneity-aware scheduling algorithm to conduct the concurrent execution of multiple DNN models with maximum usage of hardware resources. The proposed algorithm uses a greedy scheduling scheme with layer/sub-layer granularity by estimating the computation and external memory access time.
\item {\textbf{Systolic-Vector Architecture Exploration:}} We build an architecture simulation framework based on actual synthesis and place-and-route implementation results to reflect physical characteristics such as clock rate, power consumption, and area. Using this framework, we conduct design space exploration (DSE) of the proposed architecture and provide important insights and data points for advanced ML accelerator design.
\end{itemize}
With promising results based on post-layout simulation, we believe that the proposed scalable heterogeneous architecture with runtime resource scheduling is an effective and practical accelerator solution to serve ML workloads in datacenters.

%% file: Outline/2_Background_and_Motivation.tex
\section{Background and Motivation}
\label{back_motive}
\subsection{DNN Workloads in Datacenters}
As traditional AI applications, including computer vision, speech recognition, and natural language processing, use ML technology, cloud datacenters\cite{fowers2018configurable}, \cite{zheng2020optimizing}, \cite{jouppi2021ten} receive numerous DNN requests from users. Some emerging applications such as image analysis, mixed reality\cite{portal}, \cite{oculus}, \cite{wu2019machine}, and metaverse leverage multiple DNN to perform essential subtasks such as object recognition, image segmentation, target tracking, etc. From the perspective of a datacenter, the server receives a number of DNN requests that contain different model types with various sizes. Among them, convolutional neural network (CNN) based models and transformer-based models are the most popular, covering most of the ML workloads such as \cite{bochkovskiy2020yolov4}, \cite{gulati2020conformer}, and \cite{lewis2019bart}. However, each type has different workload characteristics with different basic operations.

\textbf{CNN Models}
The CNN models usually used for image-based AI applications are constructed by convolutional, pooling, and fully-connected layers. The convolutional layer scans input feature maps with multiple kernel filters by performing convolution operations (inner product between two local 3-d regions) to generate output feature maps. As both input feature maps and kernel filters are fully reused by each other, the convolutional layer raises computation bottleneck characteristics.
Then, the pooling layer is applied to reduce the size of the output feature maps along the spatial dimensions. Fully-connected layers, also called a classifier, are followed after the multiple stacks of convolutional and pooling layers to compute the scores of each class. Unlike convolutional layers, they raise memory bottleneck characteristics since weight parameters cannot be reused during matrix-vector multiplication.

\textbf{Transformer-based Models}
There are two types of models in natural language processing: discriminative and generative. Discriminative models summarize the input information and make predictions (e.g., text classification\cite{devlin2018bert}), while generative models generates new tokens after predictions (e.g., text generation\cite{lewis2019bart} and machine translation\cite{vaswani2017attention}).
Most natural language processing models are based on the transformer\cite{vaswani2017attention} structure, which consists of attention, fully-connected, and non-linear activation layers. First, the fully-connected layer generates the query, key, and value matrices, which are inputted to the attention layer. The attention layer is the main part of the transformer block, which uses the input matrices to obtain the attention matrix. There are three computational steps in the attention layer: 1) the layer multiplies the query matrix with the transposed key matrix, 2) the softmax function is applied to get the attention probabilities, and 3) the attention matrix is acquired by multiplying the attention probabilities with the value matrix.
The layer multiplies the query matrix with transposed key matrix and then applies a softmax function to get the attention probabilities. Then by multiplying the attention probabilities with the value matrix, the attention matrix is acquired. After the attention layer, the final result of the transformer block can obtain through a feed-forward network, which consists of several fully-connected layers and a non-linear activation layer. Most of the operations included in the transformer block are memory-bound operations, which have low data reusability.

\begin{figure}[t]
    \centering
    \includegraphics[width=0.45\textwidth]{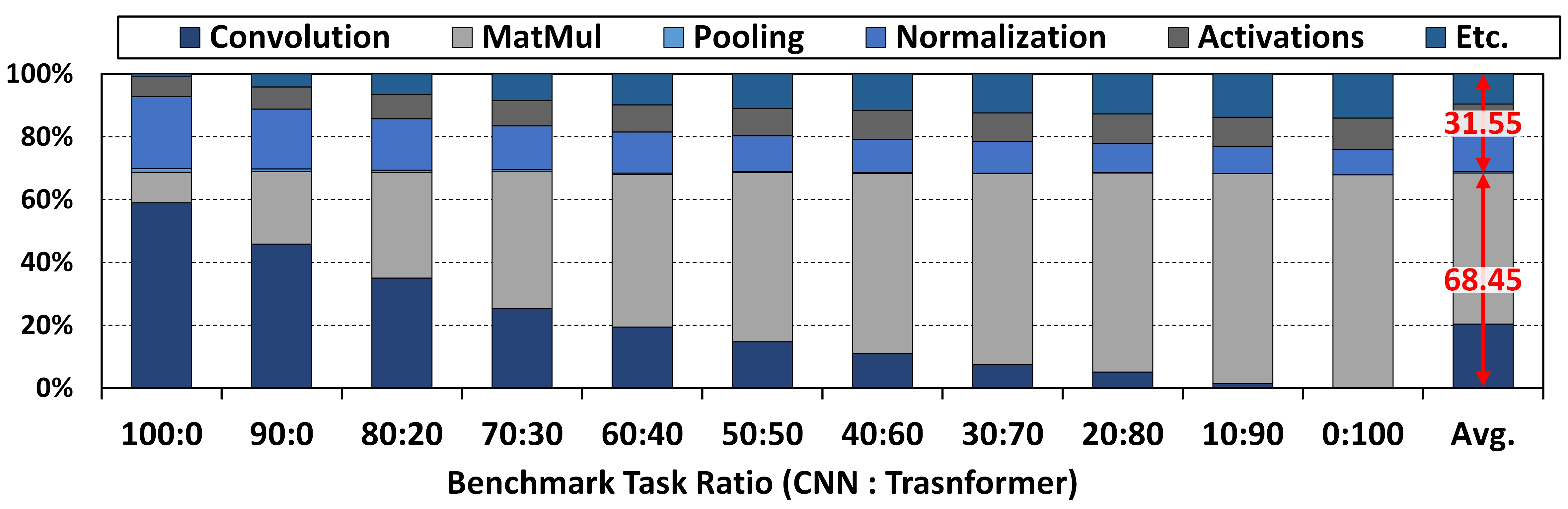}
    \caption{Breakdown of execution time for different ML operations}
    \label{fig_profile}
\end{figure}

\subsection{DNN Model Description Format}
ML developers use software frameworks such as TensorFlow\cite{tensorflow2015-whitepaper}, PyTorch\cite{NEURIPS2019_9015}, and MxNet\cite{chen2015mxnet} because they provide high-level programming abstraction and pre-implemented computation kernels. However, the conversion of models from one framework to another has been difficult because each framework has its own format to describe the DNN models. In order to solve this interoperability problem among ML frameworks, open neural network exchange (ONNX)\cite{bai2019} is developed by Microsoft and Facebook. ONNX is an open format to represent both deep learning and traditional ML models with a common intermediate representation. Specifically, ONNX uses a serializing structured data protocol called Protobuf\cite{Varda2008} that has the flexibility to describe DNN models. With the development of ONNX, ML developers can transfer any DNN model between frameworks via the shared format regardless of the framework discrepancies. It also opens up opportunities for hardware developers by targeting only ONNX as the hardware's software interface.

However, ONNX cannot be directly applied in an AI hardware accelerator due to few underlying problems. One problem is that the Protobuf protocol in ONNX has significant data redundancy to support dynamic binding. Dynamic binding is a compilation method in which a name associated with a particular operation is looked up at run-time, in this case, to resolve the differences between frameworks and thus enable flexibility. Therefore, each operation is prefixed by a name instead of the operations being grouped in a compact manner. Decoding such a robust data format causes overhead and reduces efficiency in hardware.
Another problem is that the ONNX format does not contain any user description field. As multiple users make multiple requests in the datacenters, another level of abstraction is required to describe user information when handling the model requests.
Due to these shortcomings, ONNX is hard to use in the custom AI accelerator directly.

\subsection{Scheduling Algorithms for DNN Workloads}
When executing multiple requests of DNN workload, the hardware underutilization is the main issue as there are always mismatches between the requested model requirements and the hardware's native capability. To address this problem, several scheduling algorithms have been proposed including PREMA\cite{choi2020prema}, AI-MT\cite{baek2020multi}, and Layerweaver\cite{oh2021layerweaver}.

PREMA proposes a preemptible neural processing unit (NPU) for DNN workloads. Its preemptive scheduling algorithm meets the different latency demands based on priority while maintaining the throughput in the NPU. Although PREMA improves the quality of service by complying with the restricted latency requirements, it does not improve the overall throughput because it fails to obtain the optimal scheduling for the execution of multiple DNN. AI-MT introduces a sub-layer scheduling algorithm for improving the throughput of DNN workloads using load balancing between the computation and memory access. However, it requires user-defined threshold values that are critical to scheduling performance\cite{oh2021layerweaver}. As these parameters should be obtained by the heuristic methods, suboptimal results are achieved. Lastly, Layerweaver proposes a time-multiplexing scheduling algorithm by interweaving layer execution for DNN workloads.
Layerweaver optimizes the utilization of computation and memory resources for workloads in which the ratio between the memory- and the compute-intensive task is equal. However, its advantages are minimized when the ratio is uneven because the layer cannot be interweaved. This uneven case is likely to occur in a real data center environment where the server randomly receives a number of DNN requests, including different model types with various characteristics.

\subsection{Motivation for Heterogeneous Architecture with Resource Scheduling}
DNN model is composed of two types of operations: array and vector. Array operations include convolution and matrix-matrix multiplication, both consisting of multiply-and-accumulate (MAC), and vector operations include pooling, normalization, and non-linear activation functions.
Earlier NPU architectures are designed only with a homogeneous array processor because array operations predominate the DNN models. When we profile the GPU's execution time of each operation by varying the portion of CNN and transformer models, we find that vector operations are responsible for 31.55\% of the total execution time as shown in Figure~\ref{fig_profile}. This ratio indicates that the vector operations significantly affect latency, which suggests that a heterogeneous architecture with the support of both vector and array operations would be the most effective in accelerating ML workloads. For this reason, recent NPU architectures like the TPU \cite{jouppi2021ten} include a vector-type processor along with an array-type processor. However, the vector processor only supports limited operations to assist the array processor; it only performs pooling, normalization, and non-linear activation functions. In addition, it suffers from a long idle time since it waits for the array processor to complete its task. To improve the utilization, a vector processor that can flexibly support array operation with low hardware overhead is necessary. Moreover, the pre-existing scheduling algorithms only focus on scheduling array operations within a single NPU core. A scheduling algorithm for DNN workloads that improves the utilization on both vector and array processors and among multiple cores is required for maximum performance and utilization.

%% file: Outline/3_Unified_Model_Format.tex
\begin{figure}[t] 
    \centering
    \includegraphics[width=0.45\textwidth]{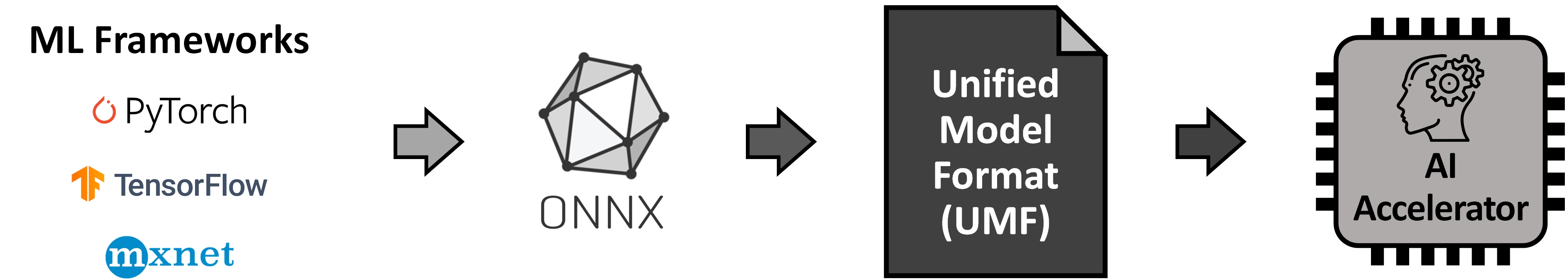}
    \caption{Unified model format overview}
    \label{fig_umf_overview}
\end{figure}

\begin{figure}[t]
    \centering
    \includegraphics[width=0.45\textwidth]{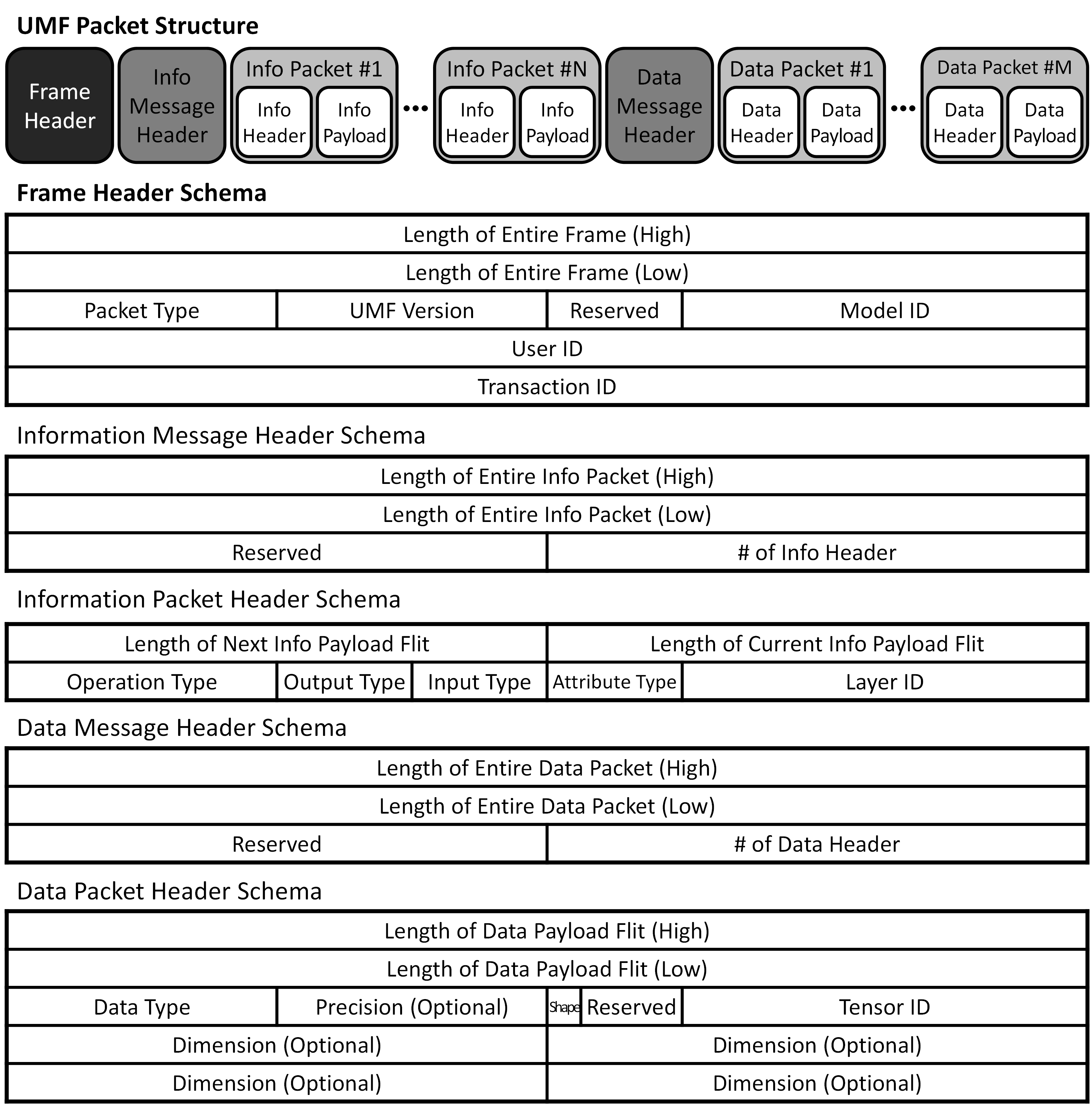}
    \caption{UMF Packet Structure and Schema}
    \label{fig_umf_structure_schema}
\end{figure}

\section{Unified Model Format}
ONNX is an efficient file format to describe DNN models, but it has compatibility issues when applied to hardware because it is designed for software frameworks as discussed in Section \ref{back_motive}. In particular, we discover problems with data redundancy and lack of user information that hinder the support of multiple DNN workloads in hardware. To amend these problems, we propose a unified model format (UMF), a lightweight DNN model format that is optimized for hardware use. UMF extracts only essential data from ONNX and packs them in a compact packet format to enable fast hardware decoding. We develop a tool that converts ONNX files into UMF binary files. Figure~\ref{fig_umf_overview} shows how DNN models enter an AI accelerator as UMF files. First, we convert DNN models from various frameworks such as PyTorch or Tensorflow into the ONNX format. Then, we decode the ONNX files and convert them into the UMF format. The UMF files are sent to an AI accelerator for model inference.

\subsection{UMF Packet Structure}
Figure~\ref{fig_umf_structure_schema} shows the packet structure of UMF. The UMF frame is composed of three stacks: a frame header, an information message header, and a group of information packets, and a data message header and a group of data packets. The frame header describes the UMF properties (e.g., packet type and UMF version) and information about the user, transaction, and DNN model. The AI accelerator interprets the UMF properties and decides how to decode the UMF packets. In addition, it obtains the information about the user, transaction, and model through the IDs included in the frame header. Based on this information, the accelerator can identify a specific request among many other in-flight requests on the chip.

The information message header indicates how many information packets exist. Each information packet contains complete information to describe a single operation layer. The information packet is composed of the header and payload. The information packet header includes the current and next payload size, layer ID, operation type, input/output type, and attribute type. The current and next payload size determines how much data to enqueue to the task queue in the accelerator. The operation type describes the layer's operator, which is either a compute operation such as Conv, GEMM, and MatMul or a data operation such as reshaping and concatenation. The input type describes the number of input tensors and each input's type (i.e., weight or activation), while the output type only includes the number of output tensors. The attribute type indicates which attributes (e.g., kernel size and padding type) and stride will be included in the information payload. A combination of input/output tensor and attribute information describes all the information necessary for operation.

Similar to the information message header, the data message header indicates how many data packets exist. Following the data message header, data packets contain the parameters of the target DNN model. Each packet header includes tensor ID, data type, precision, and size of the payload. The tensor ID is a unique ID in the DNN model that is used to index into the data packet payload that stores the model parameters. It is referenced by the information packet payload's operation information to get the tensor values used in the given operation. By decoding the data type and precision, the accelerator decides how to interpret the data in the data packet payload.

\begin{scriptsize}
\begin{figure*}[h]
    \centering
    \includegraphics[width=0.93\textwidth]{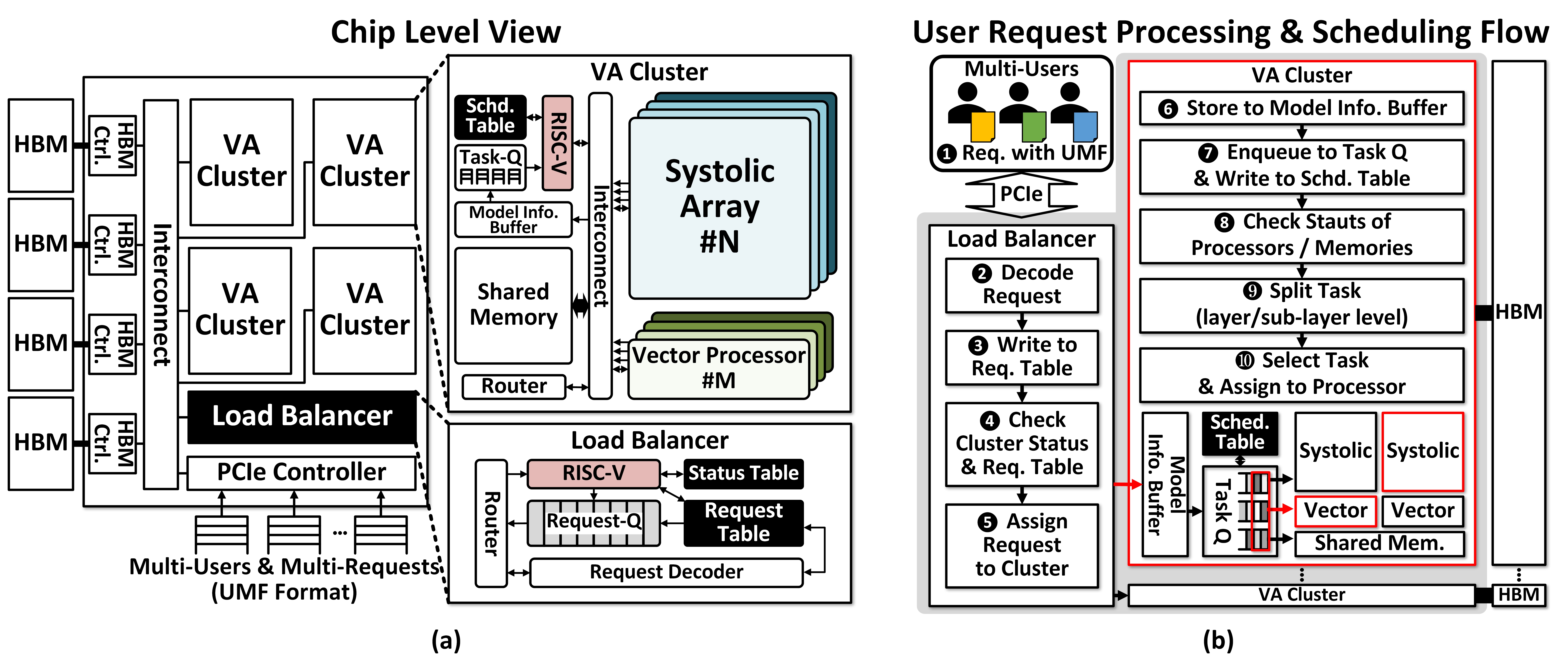}
    \caption{Heterogeneous Systolic-Vector (HSV) architecture (a) Block diagram (b) User request processing and scheduling flow}
    \label{fig_arch}
\end{figure*}
\end{scriptsize}

\subsection{UMF Packet Types}
Based on the proposed packet structure, UMF supports three packet types for different usages: \texttt{model load}, \texttt{request-return}, and \texttt{check-ack}. The \texttt{model load} type of UMF consists of a frame header, information packets, and data packets. This type is used when a user loads a DNN model to the AI accelerator. Graph information and parameters of the DNN model are converted into information and data packets, respectively, and packed into UMF with the frame header. \texttt{request-return} type of UMF is composed of a frame header and data packets. This type is used when requesting DNN model inference or when returning result values of inference request. \texttt{check-ack} type of UMF consists of only a frame header. It is used for acknowledgment when the model requested by the user is successfully loaded into the AI accelerator, when the model id is returned to the user, or when the user checks whether there is a model loaded with the given model id in the AI accelerator.

%% file: Outline/4_Heterogeneous_Vector_Array_Architecture.tex
\section{Heterogeneous Systolic-Vector Architecture}
\label{overall_arch}
Figure~\ref{fig_arch}(a) shows the overall architecture of the proposed heterogeneous systolic-vector (HSV) accelerator, consisting of a load balancer, systolic-vector (SV) clusters, high bandwidth memory (HBM) controllers, and on-chip interconnection.
In this architecture, the host CPU encodes the incoming ML inference requests from multiple users into UMF packets and sends them to the accelerator via PCI Express (PCIe). The load balancer then decodes the UMF packets and performs a high-level workload distribution to available SV clusters. Each SV cluster independently executes the assigned ML inference requests and signals back to the load balancer when it completes any one of the requests. The load balancer and SV clusters can access the external HBMs through fully connected interconnection. These components of the HSV architecture are designed to follow the user request processing flow and provide efficient hardware for multi-user requests of multiple DNN ML workload in datacenter.

\subsection{User Request Processing Flow}
Figure~\ref{fig_arch}(b) shows the overall processing flow of a user request in the HSV accelerator. \circnum{1} The host CPU receives various ML inference workloads from users and sends them to the accelerator in UMF through PCIe. \circnum{2} The UMF decoder in the load balancer decodes the UMF header to get the user description, and then \circnum{3} writes the model to the request table. \circnum{4} By checking each SV cluster status and request table, \circnum{5} the load balancer assigns the ML inference request to an available SV cluster and updates the status table. Once the assigned request enters the SV cluster, \circnum{6} it is interpreted to layer-wise tasks and stored in the model information buffer. This buffer stores the ML model information used to estimate the execution time, external memory access, and data size to fetch. \circnum{7} Then layer-wise tasks enqueue to the task queue and write their information (e.g., estimated value, transaction ID, and etc.) to the scheduling table. This table includes the information for scheduling, such as layer dependency and the status of each processor. \circnum{8} By checking the status and available resources of the underlying processors and memories, \circnum{9} the scheduler in the SV cluster may split the request into sub-layer tasks. \circnum{10} These tasks in the task queues are selected and allocated to the processors to maximize the utilization of computation resources and external memory bandwidth.

\begin{scriptsize}
\begin{figure*}[h]
    \centering
    \includegraphics[width=0.93\textwidth]{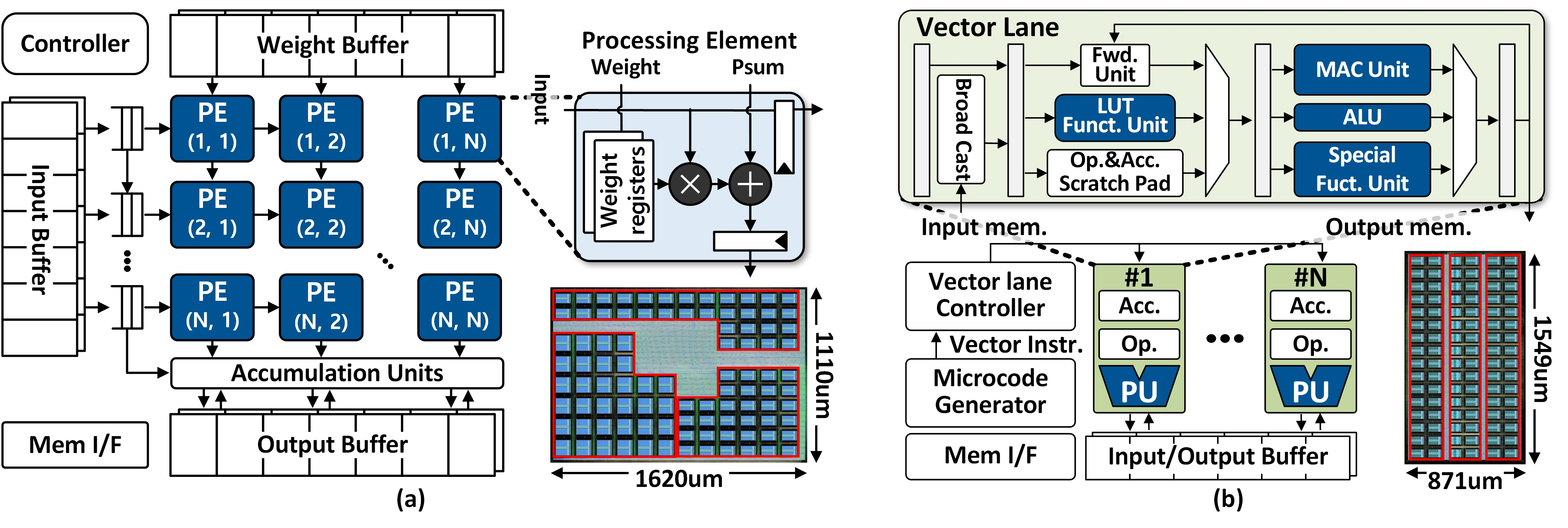}
    \caption{Processor design and layout result using 28nm standard cell library (a) Systolic array (b) Vector processor}
    \label{fig_design}
\end{figure*}
\end{scriptsize}

\subsection{Load Balancer}
The load balancer is the entry module in the accelerator that receives the user requests from the host. It consists of a UMF decoder, RISC-V controller, request queue, request table, and status table.
When the ML inference request is entered into the load balancer, the UMF decoder decodes the UMF packet to identify the user information. After that, the load balancer stores the user ID and target DNN model information in the request table. By checking the status table, the RISC-V controller allocates a new request to a SV cluster through the request queue with the first-in-first-out strategy.

\subsection{Systolic-Vector Cluster}
\label{vacluster}
SV cluster is the main operation module of the HSV architecture that executes multiple ML requests assigned by the load balancer. It consists of a RISC-V scheduler with a scheduling table and task queues, model info buffer, systolic arrays, vector processors, and shared memories. Once a request is enqueued in the task queue, the RISC-V scheduler, the central controller of the cluster, decodes the payload of the request and calculates the estimated time for computation and external memory access. Considering the performance characteristics of the underlying sytolic array, vector processors, and the capacity of shared memories, the scheduler further splits the layer-wise task into sub-layer tasks. The scheduler then assigns the sub-layer tasks to the available processors one by one. For runtime task scheduling, it manages necessary information in the scheduling table, such as a candidate task group, currently running tasks, the status of each processor, and the available capacity of the shared memories. This scheduling is a complicated problem that computes an optimal distribution of arbitrary tasks to available resources at the given time. We use a fully programmable RISC-V processor for the scheduler because there can be many different scheduling algorithms. We can apply other scheduling algorithms by updating the processor's program, even at runtime, to adapt dynamically changing workload characteristics of the datacenter to improve the performance. In the cluster, the systolic array and vector processors execute tasks assigned by the RISC-V scheduler. In addition, shared memory is designed to share the activations, parameters, or intermediate data among the processors.

\textbf{Systolic Array}
Our systolic array has a typical 2-D systolic design~\cite{jouppi2017datacenter} to enable high-throughput matrix-matrix multiplications and 3-D convolution operations. As shown in Figure~\ref{fig_design} (a), it consists of a controller, 2-D array of processing elements (PEs), and on-chip buffers for input, weight, and output data. The controller is responsible for the overall data movement in the processor, generating addresses for the input and weight buffer to feed data into the PE array.
We use the weight stationary dataflow, which preloads the weights from the weight buffer to the PEs. For matrix-matrix multiplication, each row of the weight matrix is mapped to each column of the PE array. For 3-D convolution operation, each 3-D weight kernel is flattened and mapped to each column of the PE array. With this weight mapping, the elements of an input vector are fed from left across the multiple rows of the PE array. From the top row, each element is entered with a single cycle delay. In each PE, the incoming input is multiplied by the stored weight and added with a partial sum from the above PE. With the single-cycle delay, the partial sums are constantly accumulated to the bottom of the array. The accumulation units store the intermediate partial sums from the array and accumulate them through multiple iterations for large matrix/convolution operations. The final results are stored to the output buffer.
For high utilization of the PE array, we use double buffering for both on-chip memories and internal registers of PEs. The input buffer prefetches the next input data, and output buffer writes the previous results to keep the PE array operating without stalling. Likewise, each PE loads the subsequent weight as it processes the current weight. By alternating the read registers, it can seamlessly utilize the MAC unit.

We implement the systolic array in a 28nm CMOS process. For a 16$\times$16 PE array with a pair of 16$\times$2KB input/weight and 16$\times$4KB output buffers, the systolic array takes 1.62mm$\times$1.11mm die area, achieving 800MHz operating frequency after post-place-and-route simulation.

\textbf{Vector Processor}
Figure~\ref{fig_design} (b) shows the block diagram and layout of the vector processor. It is an in-order single instruction multiple data (SIMD) processor with multiple vector lanes. It is primarily designed for running non-matrix operations in DNN models, such as pooling, nonlinear activation, and element-wise vector operations. Since the vector processor can also run matrix-matrix multiplication or 3-D convolution operation through programs, it provides more options in task scheduling.
The vector processor is composed of a microcode generator, vector lane controller, multiple vector lanes, and input/output buffers for direct memory access (DMA). Receiving a layer or sub-layer task from the scheduler in the cluster, the microcode generator generates low-level binary instructions optimized for the vector processing, alleviating programming overhead and instruction fetch cycles. It feeds the generated instructions to the vector lane controller that directly controls the cycle-level operations of the multiple lanes. The vector lane controller generates control signals for the arithmetic units, detects data hazards, and handles multi-cycle operations. If a stall event is detected, it sends the hold signal to the microcode generator. Each vector lane includes an operand/accumulation scratchpad, customized datapath, and computation units (i.e.,  MAC, ALU, special function unit, and LUT function unit) to support various ML computations. Special function unit includes reciprocal and exponent units, which take multiple cycles to operate. Combining these operations, the vector processor can perform complicated operations such as softmax function. Meanwhile, LUT function unit supports the computation of nonlinear activation functions. It selects a weight and a bias from preloaded datasets using an input value to perform the linear interpolation. After selecting the operands (e.g., input, weight, and bias), the linear interpolation output is obtained by performing the multiply-and-addition in the MAC unit. We also use double buffering for input/output buffer to hide the data fetch and write back latency in the DMA.

Like the systolic array, we implement the vector processor in a 28nm CMOS process. For 16 vector lanes with a pair of 16$\times$2KB input/output buffers, it occupies 1.55mm$\times$0.87mm die area, achieving 800~MHz operating frequency after post-place-and-route.

\textbf{Shared Memory}
Shared memory is designed to store the weight or activation data that should be shared among the processors. It is byte-addressable and has multiple banks for high throughput data communication. Each processor can concurrently access the shared memory in the cluster through the fully connected interconnection to read or write data. Utilizing shared memories, we can reduce external memory access. For example, a processor can directly use the output activations in shared memory as the input of the next layer. We can also decrease the external memory bandwidth requirement and increase data reuse by sharing the weights between tasks and between different requests using the same DNN model.

%% file: Outline/5_Scheduling_Methods.tex
\section{Scheduling Algorithms}
\label{sched_algo}
The scheduling algorithm is critical in the HSV architecture because it directly affects the utilization and overall performance of architecture. So, we develop a novel scheduling algorithm that allocates incoming tasks efficiently to the available hardware resources such as the systolic array, vector processors, on-chip shared memory, and external memories. In this section, we describe the scheduling algorithm in the HSV architecture.

\subsection{Round-Robin Scheduling}
We choose round-robin (RR) as our baseline scheduling algorithm. In the RR algorithm, the scheduler chooses a task out of a task queue in a circular order and assigns it to an available processor, either systolic array or vector processor. Even if the vector processor can perform the array operations, each type of task (e.g., array and vector operation) is only assigned to the dedicated processor. It moves to the next task queue and waits for an available resource to allocate the new task. As the scheduler iterates through the task queues in turn, underutilization is likely to occur in the accelerator as the scheduling does not consider any characteristics of tasks and processors, and their relations.

\RestyleAlgo{ruled}
\setlength{\textfloatsep}{0pt}% Remove \textfloatsep
\begin{algorithm}[t]
\caption{Heterogeneity-Aware Scheduling}
\label{algorithm_comp}
\scriptsize

\SetKwInOut{InOut}{Input/Output}
\InOut{Scheduling Table $S$, Candidate Task Group $G$}
\For{$q=0$; $q<NUM\_TQ$; $q=q+1$}{
    $t\_mem[q] \leftarrow$ extMemAccessSche($S, G[q]$) \\
    \For{$p\in [vp, ap]$}{
        $t\_task, t\_proc \leftarrow$ access $S$ \\
        $t\_start[p] \leftarrow$ max($t\_mem[q], t\_task, t\_proc$) \\
        $t\_comp[p] \leftarrow$ calcCompTime($G[q], p$) \\
        $t\_end[p] \leftarrow t\_start[p] + t\_comp[p]$
    }
    $p \leftarrow$ nominateProcessor($S, t\_end$) \\
    $t\_idle[q] \leftarrow$ calcIdleTime($S, t\_start, p$)
}
$q\_sel \leftarrow$ selectTask($S, t\_idle$) \\
scheduleAndUpdate($S, G, t\_mem, q\_sel$) \\
\end{algorithm}

\subsection{Heterogeneity-Aware Scheduling Algorithm}
We propose a heterogeneity-aware scheduling (HAS) algorithm that maximizes the overall resource utilization by minimizing the idle time of each processor based on execution time estimation. HAS algorithm has two steps. First, the scheduler partitions a layer-wise task into multiple sub-layer tasks in a way to minimize the unnecessary data movement among the processors, using the DNN model information and the hardware configuration. The model information contains parameters and activation size in each layer, and the hardware configuration refers to the numbers and sizes of systolic array and vector processor as well as the size of shared memory. Second, the scheduler assigns the multiple sub-layer tasks without dependencies to multiple processors in parallel to minimize the execution time latency. In addition, we utilize the scheduling algorithm for external memory access to maximize the performance gain using HAS. Moreover, we illustrate a simple example of HAS by comparing it with the RR algorithm.

\textbf{Detail of HAS Algorithm}
Algorithm~\ref{algorithm_comp} shows the HAS algorithm, which finds optimal scheduling for the given candidate task group $G$. The candidate task group $G$ is a group of unscheduled tasks from the multiple task queues. Each candidate task includes information about its operation and attributes as well as partitioning and dependency information if the task is sub-divided. Scheduling table $S$ records the start/end time of the assigned task for each compute resource (i.e., systolic array or vector processor) and the time when the parameters and activations required for running the task are ready.
HAS algorithm starts by estimating the execution time for each task in $T$, which adds the start time and the computation time. The scheduler first estimates the start time by checking the status of the memory, tasks, and processors. Using external memory access scheduling, which will be explained later, the scheduler obtains the memory ready time ($t\_mem$) when the required parameters and activations for the given task are ready in on-chip memory. It also accesses the scheduler table $S$ to get the end time of the dependent task for each candidate task ($t\_task$) and the earliest available time for each processor ($t\_proc$).
Once it knows $t\_mem$, $t\_task$, and $t\_proc$, the start time ($t\_start$) is the maximum time between them. Therefore, $t\_start$ represents the earliest time when the memory, tasks, and processors are all available. Then, the scheduler uses the performance model to estimate the computation time ($t\_comp$) of each processor if it runs the given task. By adding the estimated computation time to the start time, it obtains the end time ($t\_end$). Among the estimated end times from all processors, the scheduler nominates the processor ($p$) that has the earliest end time. Finally, it calculates the idle time ($t\_idle$) of the nominated processor by subtracting the end time of the previous task from the start time of the new task. HAS algorithm repeats the above process to obtain all the idle times of the candidate task group $G$. Among them, the scheduler selects the task ($q\_sel$) that has the shortest idle time for the final scheduling. If two or more scenarios have the same minimum idle time, it selects the task from the queue that is next in turn, as in RR. Once the selected task is scheduled to the processor, it updates the scheduling table $S$ and candidate task group $G$ for the next iteration.

\RestyleAlgo{ruled}
\setlength{\textfloatsep}{0pt}
\begin{algorithm}[t]
\caption{External Memory Access Scheduling}
\label{algo_ema_sche}
\scriptsize

\SetKwInOut{Input}{Input}
\SetKwInOut{Output}{Output}
\Input{Scheduling Table $S$, Candidate Task $T$}
\Output{Activations\&Parameters-Ready Time $t$}
$p\_size \leftarrow$ getParamSize($T$) \\
$a\_size \leftarrow$ getActSize($T$) \\
$t\_mem \leftarrow$ getLastMemFetchEndTime($S$) \\

\uIf{\bf{parameters exist in shared memory}}{
    $t, F \leftarrow$ getParamReadyTime($S, T$), 0
}
\uElse{
    $t, F \leftarrow t\_mem, p\_size$
}
\uIf{\bf{activations need to read from external memory}}{
    $t \leftarrow t\_mem$
}
\uIf{$t = t\_mem$}{
    $R \leftarrow SM\_SIZE$ - getUsedMemSize($S$) \\
    fetchParam($T, t, R, F$) \\
    \uIf {not finishFetchParam($F$)}{
        \For{$sched\_T \in S$}{
            \uIf{$t \leq$ getEndTime($sched\_T$)}{
                updateTime($sched\_T, t$) \\
                \uIf{\bf{activations need to write to external memory}}{
                    $t \leftarrow t$ + getActWriteTime($sched\_T$)
                }
            }
            \uIf{no task uses sched\_T's parameters}{
                flushSM($sched\_T, R$)
            }
            fetchParam($T, t, R, F$) \\
            \uIf{finishFetchParam($F$)}{
                \bf{break}
            }
        }
    }
    $t \leftarrow t$ + getActReadTime($a\_size$)
}
\end{algorithm}

\begin{figure}[t]
    \centering
    \includegraphics[width=0.45\textwidth]{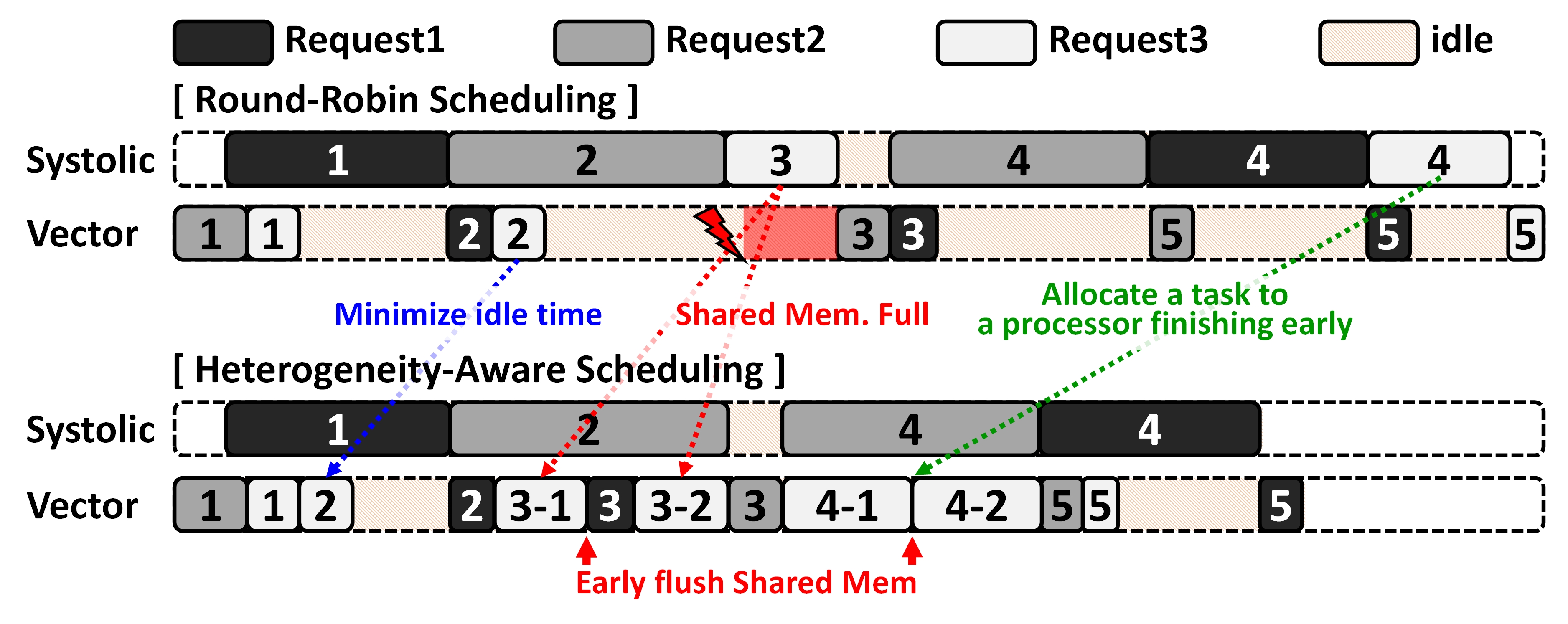}
    \caption{Example of heterogeneity-aware scheduling algorithm}
    \label{fig_scheduling_example}
\end{figure}

\textbf{External Memory Access Scheduling}
Algorithm \ref{algo_ema_sche}, the external memory access scheduling, schedules the read-and-write of parameters and activations from external memory based on the dynamic analysis of shared memory when the task arrives. To schedule the external memory access, the scheduler refers to the scheduling table $S$ and the candidate task $T$, a task that needs to be served, to calculate the ready time $t$ for parameters and activations. First, the scheduler calculates the data size of the parameters $p\_size$ and activations $a\_size$ using the information of the candidate task $T$. Then, it gets the time when it was last fetched from the external memory $t\_mem$. If required parameters are previously scheduled and exist in shared memory, the processors use the value without unnecessary external memory access. On the other hand, if the external memory access is inevitable (i.e., not already contained in shared memory), the scheduler decides how many portions of the parameter to fetch based on the remaining capacity in the shared memory. The scheduler stalls the external memory access until enough space is available. The space becomes available when the previous tasks finish and no other tasks need the given parameter. Meanwhile, when activations of the task cannot fit into the shared memory, the scheduler partitions the activations into multiple pieces. If the given activations of the sub-layer task fill the entire shared memory space, the scheduler writes the data to the external memory and flushes the shared memory to obtain enough space for other activations.

\textbf{Scheduling Example}
Figure \ref{fig_scheduling_example} shows an example of a scheduling scenario in a single SV cluster. The color of the box represents each request, and the number in the box represents the order of tasks within the request. The orange patterned box represents the idle time, in which the processor does not perform any computation. The above timetable shows the scheduling result with RR, and the bottom timetable shows the scheduling result with HAS.

Compared to RR scheduling, HAS reduces the idle time of the vector processor by scheduling the second task of request 3 before the second task of request 1. The third task of request 3 has many required parameters, so the parameters of the third task of request 2 cannot be fetched, and the vector processor waits for the end of the third task of request 3 in the existing scenario. In HAS, the memory capacity requirement for each sub-task is reduced by dividing the third task of request 3 into sub-tasks. Whenever a sub-task finishes, parameters are flushed from the shared memory to reduce the on-chip memory resource overhead. By allocating the fourth task of request 3 to the vector processor instead of the systolic array, HAS reduces the computational load of the systolic array and the total computation time. With these characteristics, HAS achieves better resource utilization and faster computation time compared to the RR scheduling.

%% file: Outline/6_Evaluation_and_Result.tex
\begin{scriptsize}
\begin{figure}[t] 
    \centering
    \includegraphics[width=0.48\textwidth]{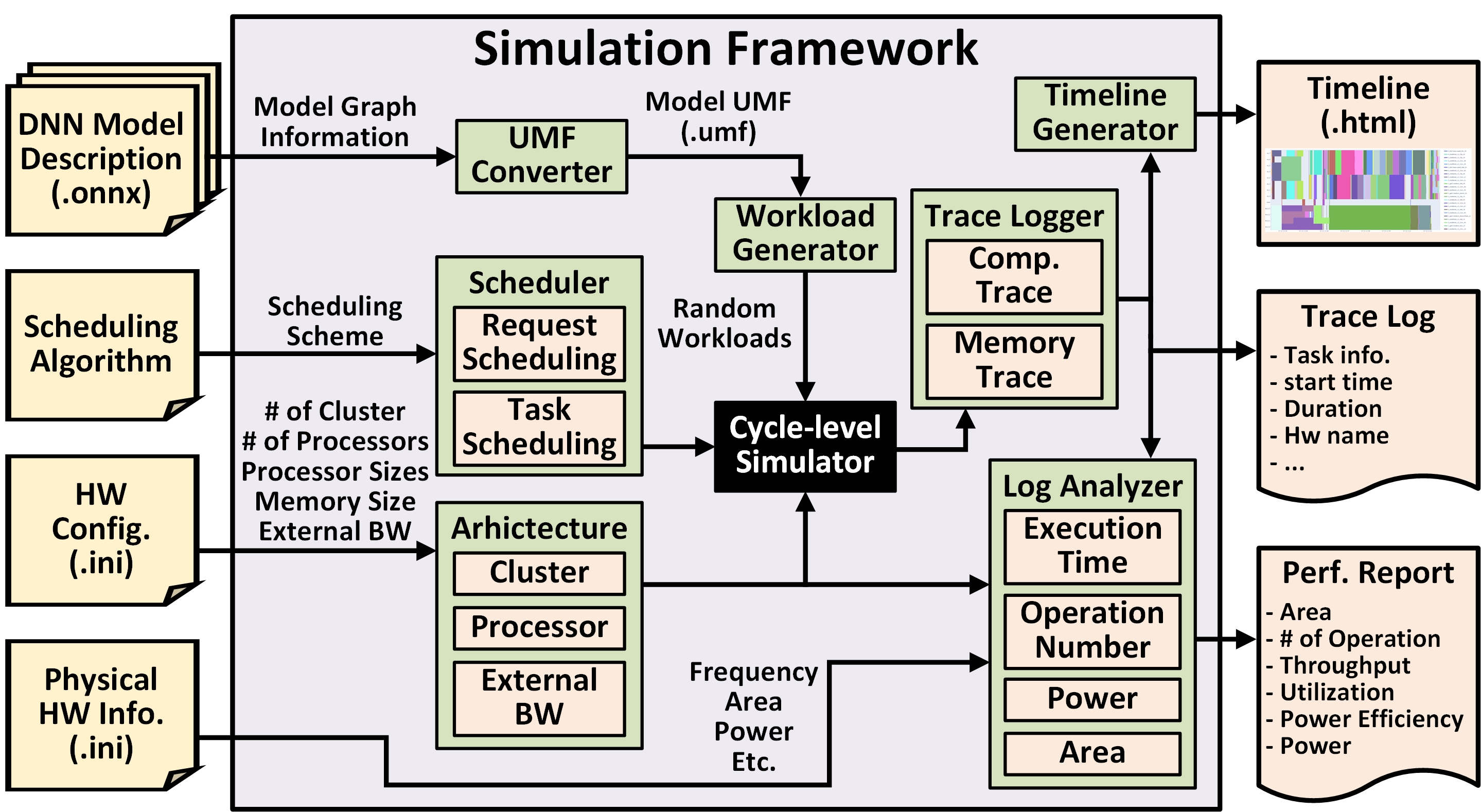}
    \caption{Simulation framework overview}
    \label{fig_framework}
\end{figure}
\end{scriptsize}

\begin{scriptsize}
\begin{figure*}[t] 
    \centering
    \includegraphics[width=0.95\textwidth]{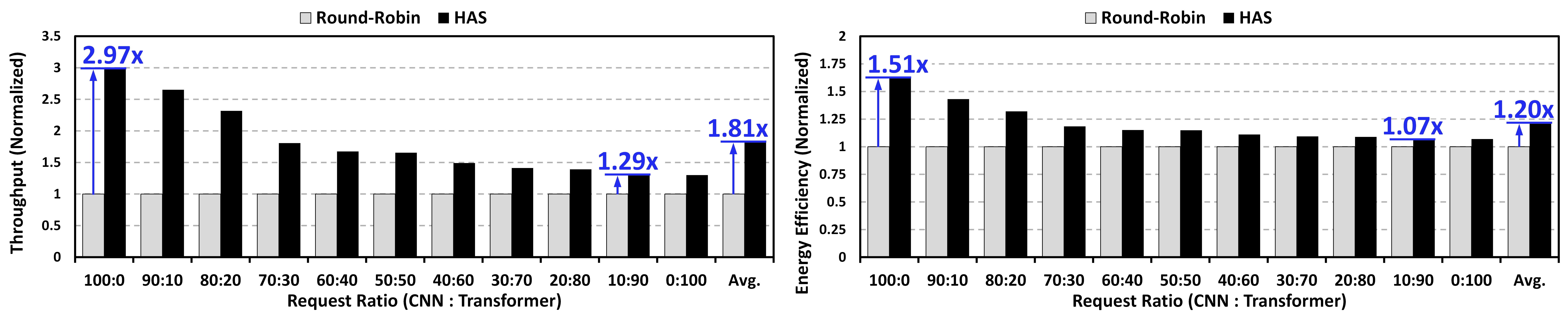}
    \caption{Throughput performance and energy efficiency improvement by heterogeneity-aware scheduling over round-robin}
    \label{algo_comp}
\end{figure*}  
\end{scriptsize}

\begin{table}
\renewcommand{\arraystretch}{1.3}
\caption{Physical Specification of Each Processor}
\label{table_spec}
{\scriptsize
\centering
\begin{adjustbox}{width=0.48\textwidth}
\begin{tabular}{c|c|c|c|c|c|c|c} 
\hline\hline
\multicolumn{2}{c|}{}                                                                                                                           & \multicolumn{3}{c|}{\textbf{Vector Processor}} & \multicolumn{3}{c}{\textbf{Systolic Array}}  \\ 
\hline
\multicolumn{2}{c|}{\textbf{Dimension}}                                                                                                         & 16      & 32      & 64               & 16$\times$16    & 32$\times$32    & 64$\times$64                             \\ 
\hline
\multicolumn{2}{c|}{\textbf{Peak Perf. [GOPs]}}                                                                                                 & 25.6      & 51.2     & 102.4              & 409.6  & 1638.4  & 6553.6                          \\ 
\hline
\multicolumn{2}{c|}{\textbf{Area [mm\textsuperscript{2}]}}                                                                                      & 1.25       & 2.53       & 5.08               & 1.69  & 4.35  & 13.00                           \\ 
\hline
\multirow{6}{*}{\begin{tabular}[c]{@{}c@{}}\textbf{Energy}\\\textbf{per}\\\textbf{Operation}\\\textbf{[pJ/op]}\end{tabular}} & \textbf{MAC}     & 6.11       & 6.16       & 6.19                & 2.07  & 1.33  & 0.38                           \\ 
\cline{2-8}
                                                                                                                             & \textbf{Pooling} & 17.9 & 18.0 & 18.1          & \textbf{-}  & \textbf{-}  & \textbf{-}                           \\ 
\cline{2-8}
                                                                                                                             & \textbf{LUT}     & 21.7          & 21.9          & 22.0                   & \textbf{-}  & \textbf{-}  & \textbf{-}                           \\ 
\cline{2-8}
                                                                                                                             & \textbf{Reduction}    & 27.3          & 27.6          & 27.7                   & \textbf{-}  & \textbf{-}  & \textbf{-}                           \\ 
\cline{2-8}
                                                                                                                             & \textbf{Softmax} & 155.8 & 157.3 & 158.0          & \textbf{-} & \textbf{-} & \textbf{-}                          \\
\cline{2-8}
                                                                                                                             & \textbf{etc}  & 33.7  & 34.0  & 34.1          & \textbf{-} & \textbf{-} & \textbf{-}                           \\ 
\hline\hline
\end{tabular}
\end{adjustbox}
}
\end{table}

\section{Evaluation}
\subsection{Methodology}
\textbf{Simulation Framework}
To evaluate the proposed scalable heterogeneous architecture and scheduling algorithms, we build a simulation framework in Python, which includes the DNN model description converter, cycle-level simulator, performance analyzer, and timeline visualizer. We conduct a design space exploration using our simulation framework to figure out the optimal configuration and scheduling algorithm for DNN workloads in datacenters.

Figure~\ref{fig_framework} shows the high-level overview of our simulation framework. The simulation framework takes in four different information as its input: DNN model description in ONNX format, scheduling algorithm, hardware architecture configuration, and physical hardware information (i.e., frequency, area, and power). First, the UMF converter converts the model information from the ONNX to the UMF format. Then, the scheduling algorithm and hardware configuration files determine the scheduling scheme and hardware architecture (i.e., the number of clusters, the number and the size of processors, the size of shared memories, etc.). Then, the cycle-level simulator evaluates the architecture and algorithm and records the computation and memory trace in a log. Lastly, the framework generates the timeline visualization and performance report based on the trace log and the physical hardware information.

To evaluate our cycle-level simulator, we cross-validate it against the RTL implementations. Our cycle-level simulator achieves 99.35\% accuracy in cycle count for major ML operations including convolution, matrix multiplication, non-linear activation, and pooling. We also use a cycle-accurate DRAM simulator, DRAMsim3\cite{li2020dramsim3}, for modeling the access time and power consumption of the external memory access.

Furthermore, we synthesize and place-and-route the RTL implementation for core building blocks (i.e., systolic array, vector processor, and shared memory) using Synopsys Design Compiler\cite{designcompiler} with a 28nm standard cell library. We measure the area and energy consumption in post-layout simulation at 800MHz operating clock frequency using Synopsys PrimePower\cite{primepower}. Figure~\ref{table_spec} lists the peak performance, area, and energy consumption of systolic arrays and vector processors for three hardware configurations each, which are augmented into our simulator. We carefully extrapolate each component of the baseline implementations, i.e., 16x16 for array and 16-lane for vector, to calculate the values for 32x32/32-lane and 64x64/64-lane versions. We utilize the vendor's memory compiler for shared memory characterization.

\textbf{Workload Generation}
To emulate ML workloads in datacenters, we utilize four popular CNN models (ResNet50\cite{he2016deep}, VGG16\cite{simonyan2014very}, MobileNetV2\cite{sandler2018mobilenetv2}, AlexNet\cite{krizhevsky2012imagenet}) and four transformer models (bert-base-cased\cite{devlin2018bert}, bert-large-cased\cite{devlin2018bert}, gpt2\cite{radford2019language}, gpt2-medium\cite{radford2019language}). Realistic workloads of multi-user requests are generated by varying the portions of the CNN and transformer models. To represent the order of the requests, we attach the time information on every request before entering the scheduling process. The specific set of models are selected randomly, but the ratio between CNN and transformer workloads is chosen systematically. We increment the ratio of CNN models by 10\% from 0\% to 100\% to construct a total of 11 synthetic workloads.

\begin{scriptsize}
\begin{figure*}[t] 
    \centering
    \includegraphics[width=0.95\textwidth]{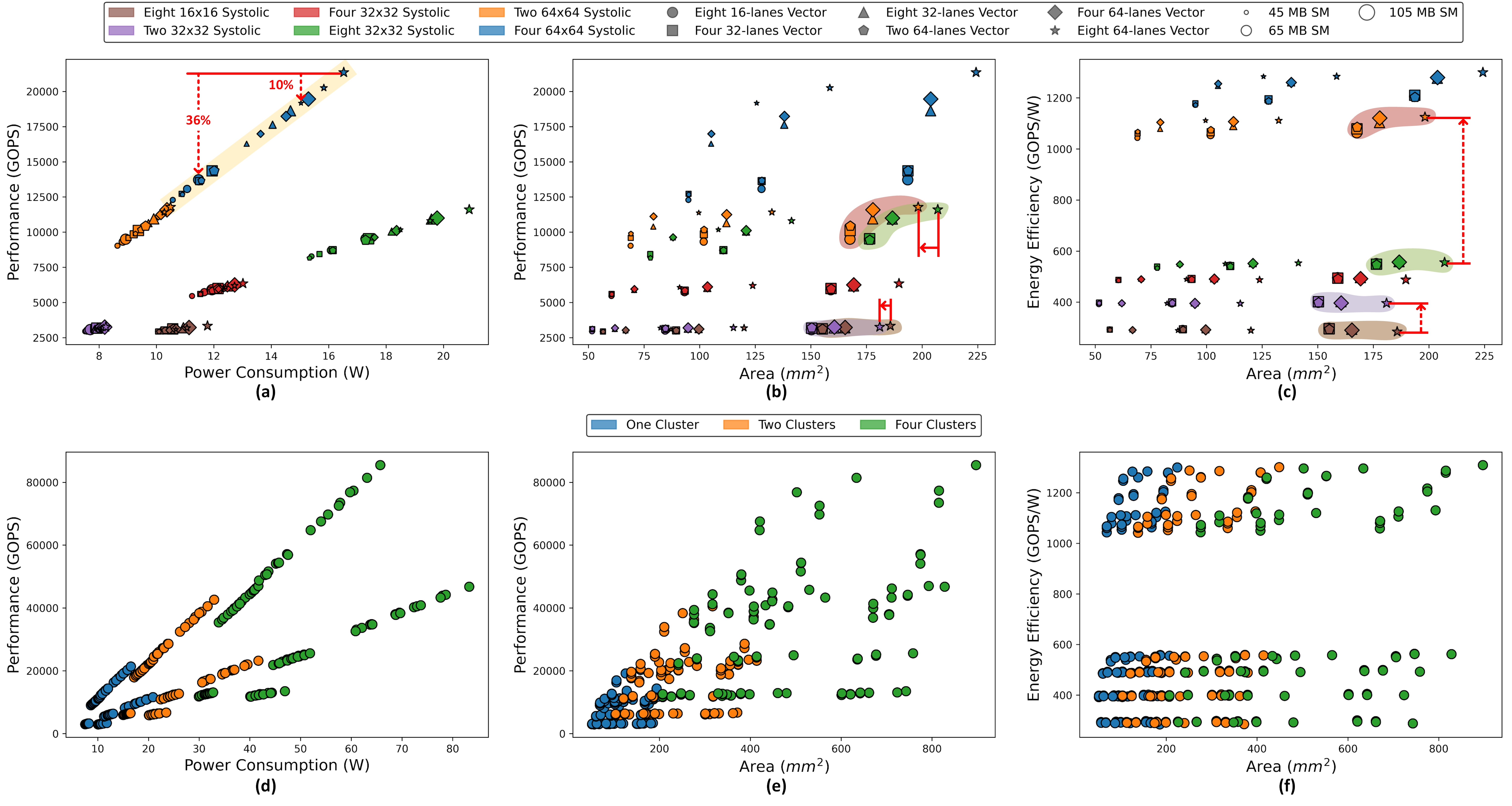}
    \caption{Design space exploration of heterogeneous systolic-vector architecture using simulation framework: (a) performance vs. power, (b) performance vs. area, and (c) energy efficiency vs. area on various architecture configurations in a single cluster and (d) performance vs. power, (e) performance vs area, and (f) energy efficiency vs. area on different number of clusters}
    \label{PPA}
\end{figure*}  
\end{scriptsize}

\subsection{Scheduling Algorithm Comparison}
To evaluate the effect of the scheduling algorithm, we measure the throughput performance and energy efficiency of various hardware configurations (i.e., different SV cluster numbers, different systolic array/vector processor numbers per cluster) on the generated ML workloads, with the two scheduling algorithms: the baseline RR and proposed HAS. Figure~\ref{algo_comp} shows the throughput and energy efficiency results of HAS normalized to the baseline results. On average, HAS shows 1.81$\times$ higher throughput than the baseline algorithm. The result indicates that the resource-aware scheduling can drastically improve the overall performance of the accelerator by improving the utilization of heterogeneous resources. One noticeable trend in the graph is that the performance gain by HAS decreases as the portion of transformer-based models increases in the workload. We analyze that this is because of the computational characteristics of the transformer-based models, which include a significant number of vector operations. This characteristic hinders HAS from allocating the array operations to the vector processors, causing HAS to exclusively assign vector operations to the vector processors, which is similar to the RR algorithm.
In addition, HAS shows 1.20$\times$ higher energy efficiency than the RR algorithm on average, with the similar trend as in performance. However, the gain in energy efficiency is reduced as the HAS consumes higher power due to high utilization. As a result, HAS improves the throughput and energy efficiency by 1.29-2.97$\times$ and 1.07-1.51$\times$, respectively, than the RR algorithm.

\subsection{Design Space Exploration}
\label{dse}
We perform the design space exploration of the proposed HSV architecture using our simulation framework. In the SV cluster, we allow six types of systolic arrays (eight 16$\times$16, two 32$\times$32, four 32$\times$32, eight 32$\times$32, two 64$\times$64, and four 64$\times$64), six types of vector processors (eight 16-lanes, four 32-lanes, eight 32-lanes, two 64-lanes, four 64-lanes, and eight 64-lanes), and three types of shared memories (45MB, 65MB, and 105MB). We measure the performance, power, and area of each hardware configuration for 33 generated workloads (3 workloads per CNN:transformer ratio). Overall, we evaluate 108 hardware configurations within the single SV cluster, resulting in 3,564 data points. Figure~\ref{PPA} (a)-(c) show the experimental results plotted in performance vs. power, performance vs. area, and energy efficiency vs. area graph, respectively. In the graphs, color and shape represent the type of systolic arrays and vector processors, respectively, while the size of markers represents the size of shared memory. We use HAS algorithm for runtime task scheduling.

\textbf{Sensitivity to Performance}
Figure~\ref{PPA} (a) shows the experimental result of the throughput performance according to power consumption on the various configurations within the SV cluster. We experiment on a single cluster first to see the impact of each processor and the size of shared memory. We discover that the throughput of the systolic arrays determines the overall accelerator's performance. Meanwhile, we also figure out that the size of vector processors and shared memories matters for performance. As shown in the graph, the throughput of the data points in the same systolic array type increases proportionally as the size and the number of vector processors and the shared memory capacity increase. We also find that the size and the number of vector processors are more critical for system performance than the shared memory size. For instance, when we change the shared memory size from 105~MB to 45~MB on the best performing array configuration, i.e., four 64x64 systolic arrays, we see 10\% decrease in throughput. However, when we chance the vector processor's lane size from 64 to 8, we see 36\% decrease in throughput. As we see the same trend for other array configurations, we conclude that optimizing vector processor configuration is more important than the shared memory capacity for higher performance.

\begin{scriptsize}
\begin{figure*}[t] 
    \centering
    \includegraphics[width=0.93\textwidth]{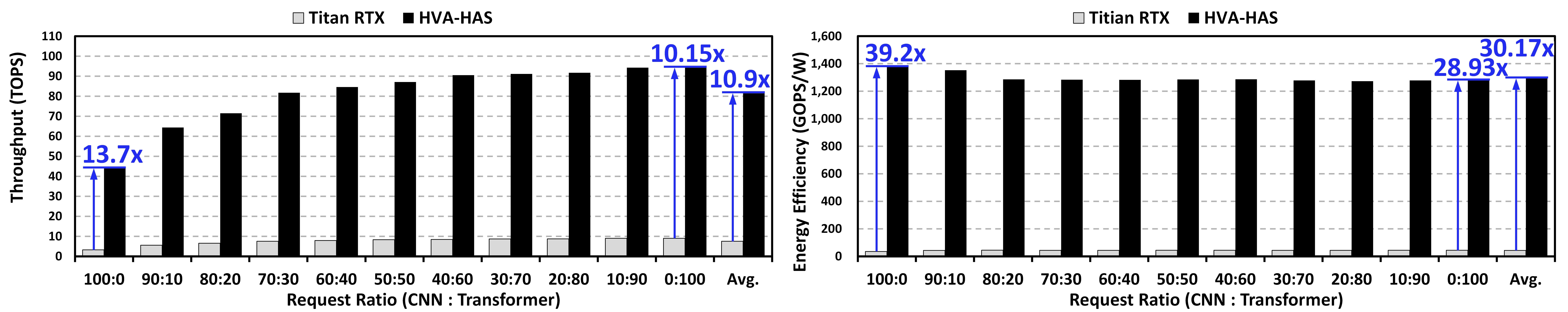}
    \caption{Throughput performance and energy efficiency improvement by the proposed heterogeneous systolic-vector architecture over Nvidia Titan RTX GPU}
    \label{gpu_comp}
\end{figure*}  
\end{scriptsize}

\textbf{Sensitivity to Efficiency}
Figure~\ref{PPA}(b) and (c) show the tendency of throughput performance and energy efficiency according to the area on various configurations within the SV cluster. As shown in both graphs, the systolic array is a dominant factor in performance and energy efficiency.
One noticeable observation is that the architectures with large-but-few systolic arrays, e.g., two 64x64 systolic arrays, have better area efficiency than small-but-many systolic arrays, e.g., eight 16x16 systolic arrays, among the architectures that yield similar performance. The main reason is that a bigger systolic array has higher energy/area efficiency than a smaller one because it has little control and buffering overhead. Another observation is that different architectural configurations with comparable number of processing elements achieve similar performances and energy efficiency. This indicates that our HAS algorithm ensures high utilization of underlying resources for any configuration.

\textbf{Cluster Scalabitity}
Figure~\ref{PPA} (d)-(f) show the experimental results plotted in performance vs. power, performance vs. area, and energy efficiency vs. area graph, when the cluster number varies among 1, 2, and 4. We run the same hardware and workload configurations as in the single cluster for multiple cluster cases. Different color represents different cluster in the graphs.
We confirm that the overall performance increases proportionally with the number of clusters. Meanwhile, the energy efficiency is maintained at similar levels even the number of clusters scales up to four. We analyze that this is because the low-overhead load balancing at the top level works well across the multiple clusters.

\subsection{GPU Comparison}
For the comparison with an existing ML accelerator, we choose Nvidia's Titan RTX GPU\cite{titanrtx} fabricated in 12nm process technology. It has a total of $754mm^2$ die area. We run the PyTorch framework\cite{NEURIPS2019_9015} with CUDA accelerated DNN libraries\cite{cuda} on the actual hardware with the boost clock frequency of 1.77GHz. Based on the design exploration results, we also choose the hardware configuration of our architecture to have a comparable die area with the GPU. Finally, we set four SV clusters, in which each cluster has four 64$\times$64 systolic arrays, eight 64-lane vector processors, and 40MB shared memories at 800MHz clock frequency, resulting in $633.8mm^2$ in 28nm process. It occupies only 84\% die area compared to the comparing GPU's even without process node scaling. 
We measure the throughput performance and energy efficiency of our heterogeneous architecture with heterogeneity-aware scheduling (HSV-HAS) and the Titan RTX GPU on the 33 synthetic workloads. 
Figure~\ref{gpu_comp} shows the throughput and energy efficiency results. HSV-HAS achieves 81.45 tera operations per second (TOPS) performance and 12.96TOPS/W energy efficiency at 800MHz operating frequency, which are 10.9$\times$ higher and 30.17$\times$ more energy efficient than the GPU on average. In particular, HSV-HAS shows higher improvements over GPU for CNN-oriented workloads. The main reason is that the systolic arrays with dedicated kernel mapping perform convolution operations more efficiently than the GPU's CUDA cores. In addition, HAS algorithm achieves optimal scheduling for the CNN models than the transformer models. In summary, the proposed heterogeneous systolic-vector architecture achieves 13.7-10.15$\times$ higher performance and 39.2-28.93$\times$ higher efficiency than a compatible GPU for various ML workloads in datacenters.

%% file: Outline/7_Related_Work.tex
\section{Related Work}

\textbf{Multiple DNN Accelerator}
HDA\cite{kwon2021heterogeneous} introduces heterogeneous dataflow accelerators, which consist of sub-accelerators that support different dataflows. Planaria\cite{ghodrati2020planaria} is the runtime composable multiple systolic arrays with mesh-based network-on-chip. It examines the workloads and connects the composable systolic arrays for each layer. These proposals only consider array processors that perform array jobs, although vector jobs have significant portions on recent workloads.

\textbf{Multiple DNN Scheduling}
To efficiently handle the multiple DNN requests in AI accelerator, multiple DNN scheduling algorithms have been studied \cite{choi2020prema, baek2020multi, oh2021layerweaver}. PREMA\cite{choi2020prema} proposes a preemptive scheduling algorithm to meet the latency requirements based on priority. AI-MT\cite{baek2020multi} devises a scheduling algorithm that balances computation and memory access to optimize throughput. Layerweaver\cite{oh2021layerweaver} introduces a multiple DNN time-multiplexing scheduling algorithm by interweaving layer execution to improve the utilization. These proposals aim to improve QoS for the latency of request or system throughput while only considering array operations. Instead, our proposed HAS demonstrates throughput improvement with array operations and vector operations by scheduling the heterogeneous systolic-vector architecture for multiple DNN workloads.

%% file: Outline/8_Conclusion.tex
\section{Conclusion}
With the analysis that ML workloads in datacenters include significant portions of vector operations, vector processors are indispensable for the ML workloads. Therefore, we present HSV, a scalable heterogeneous architecture using systolic arrays and vector processors. To achieve maximum performance of HSV, we integrate UMF, a light DNN model description, and HAS, an efficient scheduling algorithm for dynamic ML inference workloads. In addition, we build the simulation framework to perform the design space exploration for different hardware architectures and scheduling algorithms. Through the design space exploration, we gain three insights: 1) the overall performance is highly dependent on the systolic array, but the vector processors are critical to cover various ML workloads and increase the utilization of both, 2) HAS algorithm achieves high utilization of hardware resources for any configuration of systolic-vector architectures, and 3) the VA cluster supports scalability that increases performance linearly with the number of clusters. Experimental results show that the proposed scheduling algorithm can be 1.81$\times$ faster and 1.20$\times$ more efficient than a naive baseline algorithm. Also, the proposed architecture can be 10.9$\times$ faster and 30.17$\times$ more energy efficient than a comparable GPU. Showing promising results, we believe that the proposed scalable heterogeneous architecture with runtime resource scheduling is an effective and practical accelerator solution to serve ML workloads in datacenters.